\documentstyle [12pt,aaspp]{article}
\begin {document}
\title {Emergence of Filamentary Structure in Cosmological Gravitational
Clustering}
\author {B.S. Sathyaprakash and Varun Sahni}
\affil {Inter-University Centre for Astronomy \& Astrophysics,
Post Bag 4, Ganeshkhind, Pune 411007, India}
\author {Sergei F . Shandarin}
\affil {Department of Physics and Astronomy, University of Kansas,
Lawrence, KS 66045}

\begin {abstract}
The morphological nature of structures that form under gravitational
instability has been of central interest to cosmology for over two
decades. A remarkable feature of large scale structures in the
Universe is that they occupy a relatively small fraction 
of the volume and yet show
coherence on scales comparable to the survey size. 
With the aid
of a useful synthesis of percolation analysis and shape statistics
we explore the evolution of morphology of isolated density clumps
in {\it real space} and 
that of the cluster
\footnote{In this paper we call a cluster any connected region with 
density above a given threshold.}
 distribution as a whole in scale-invariant cosmological
models of gravitational instability. Our results, based on an
exhaustive statistical analysis, indicate that at finite density thresholds
one-dimensional filaments
are more abundant than two-dimensional sheets (pancakes) at most
epochs and for all spectra although the first singularities could be
pancakelike. 
Both filamentarity
and pancakeness of structures grow with time 
(in scale-free models this is equivalent to an increase in resolution)
leading to the 
development of a long coherence length scale in simulations.

\end {abstract}

\bigskip
Subject headings: Cosmology : large-scale structure of Universe -- theory.

\bigskip
{\bf To appear in Astrophys. J. Lett. 1996}
\vfill\eject

Visually structures on scales of $\sim$ 50-100 Mpc can readily be discerned
in the distribution of galaxies, 
in fact virtually all 
available catalog's reveal structures of the same size as the scale of 
the survey. A related and not completely understood issue concerns the
morphology of structures (such as the Great Wall)
observed both in the Universe and in our
numerical simulations of it.
Over a decade ago, and long before
large galaxy catalogs were available, Zel'dovich (1982) 
wondered at the
smallness of the filling factor of the galaxy distribution and Shandarin
(1983) suggested
the use of percolation theory to explore the 
topological properties of the distribution of galaxies 
(see brief summary in Shandarin \& Zel'dovich (1983)). Owing partly
to the peculiar shapes of galaxy survey volumes and partly to our
lack of understanding as to how to apply percolation analysis
under such circumstances, there has till very recently not been much
progress in this direction. (For a recent study of percolation analysis
of LSS see eg. Klypin \& Shandarin (1993), Dominik \& Shandarin (1992), 
Yess \& Shandarin (1996) and Sathyaprakash, Sahni \& Shandarin (1995).
For a review of other discriminators of large scale structure see 
Sahni \& Coles (1995).)
Although, percolation properties alone are not very useful in quantifying
the morphology of structure they do aid in its qualitative
understanding as illustrated by the following argument:
the {\it percolation transition} in an infinite volume 
characterizes the transition from 
a state in which no ``infinite''
cluster exists to one with an infinite cluster. 
For a random Poisson
distribution of points it is well known that the percolation
transition occurs at a filling factor of 0.319 (in three dimensions). 

One way of understanding this is to imagine that each particle in a distribution is surrounded by a `sphere of influence' of radius $R$. We can hop from
one particle to the next if the inter-particle separation is smaller than
$2R$. Clearly if $R$ is too small traversing the entire length of the box
in this manner is impossible. 
By increasing $R$ however, a critical value $R_c$ is
reached at which percolation occurs and it becomes possible to travel
across the entire box by jumping from one particle to the next.
In the case of a Poisson random distribution of points
the fraction of volume occupied by the union of equal spheres
centered around every particle
is about 32\% of the total volume at the percolation threshold. 
Visually a random process does not possess any distinctive
features, consequently the infinite cluster is expected to occupy a sizable
fraction of the total volume in clusters as indeed it does. 
However, particles that are
initially distributed randomly do acquire visually recognizable features
while evolving under gravitational instability, additionally the
percolation transition occurs at significantly lower filling factors
at late times.
In fact the percolating phase occupies a fraction 
of the total volume as small as 2-7\% in models such as CDM and CHDM
(\cite{ks93}). 
This immediately suggests that individual
structures within the percolating phase are unlikely to be spherical
but ought
to be more like sheets or filaments. In fact lower filling factors at 
the percolation transition
are likely to favor structures that are filamentary rather than planar
({\it i.e.} one-dimensional rather than two-dimensional), since the former 
occupy a comparatively smaller volume than the latter.

In order to quantify morphology we have to resort to statistical
indicators other than percolation. 
In recent years several such measures have been introduced.
After a careful study (\cite{sss95}) we
have chosen the structure functions of Babul and Starkman (1992)
to characterize morphology of structure at the percolation transition.
In our opinion these structure functions are quite robust and unbiased
and are consistent with the visual impression that one forms from a
given simulation. In the remainder of this Letter we describe the
percolation properties and the evolution of morphology 
of {\it the mass distribution in real space}
in
scale-invariant models evolving under gravitational instability.
A detailed report of
our study will be published elsewhere (\cite{sss95}).

The models we have studied are N-body simulations of scale invariant spectra
with power spectrum $P(k) \equiv \left <
\left | \delta_k \right |^2 \right > \propto k^n,$ for $k<k_{Ny},$ and
$P(k) = 0$ for $k\ge k_{Ny},$ $n=-3,$ $-2$, $-1$ $0$ and $+1$, where $k_{Ny}$ is
the Nyquist cutoff, $\delta_k$ is the Fourier transform of the density
contrast $\delta({\bf x})= (\rho ({\bf x}) - \overline {\rho})/\overline
{\rho},$ $\rho ({\bf x})$ being the density and $\overline \rho$ its
average, $\left< ~\right >$ denotes the average over an ensemble.
We consider several epochs $\sigma$ each characterized by the scale of
nonlinearity $k_{\rm NL}^{-1}$ at that epoch determined by
\begin {equation}
\sigma \equiv \left <\delta^2\right >^{1/2} = D_+(t)
\left (4 \pi \int_0^{k_{\rm NL}} P(k) k^2 dk \right )^{1/2} = 1
\end {equation}
where $D_+(t)$ is the linear growing-mode solution of density fluctuations.
N-body simulations are performed on a grid of size 128$^3$ using an equivalent
number of particles and employing a particle-mesh algorithm (for details see
Melott \&
Shandarin 1993).
At each epoch density fields are constructed on a reduced grid of
size 64$^3$ using a cloud-in-cell (CIC) algorithm 
\footnote {We use a simple CIC algorithm in which the mass in a particle 
is reassigned to its six nearest neighbor lattice points: if the distance to
a lattice point from the particle is $(x,y,z)$ then the mass assigned of the 
former is $(1-x)(1-y)(1-z)$.
Smearing, such as the one
we employ, amounts to smoothing the density field by removing 
fluctuations on the Nyquist frequency scale.}
and our studies of
percolation and morphology are carried out on these fields.
For the sake of brevity we present results only for
$n=-2$ and $n = 0$ models, at epochs when the nonlinear length scale is
one of the following: 
$k_{\rm NL}=64,$  16 and 4 in Fig. 1 \& 2; 
$k_{\rm NL}=32,$  16, 8 and 4 in Fig. 3;
$k_{\rm NL}$ is measured in units of the fundamental
mode $k_f\equiv 2\pi/L,$ $L$ being the length of the simulation box.
Results obtained for other models are qualitatively similar
to those reported here and will be discussed elsewhere [\cite{sss95}].

Percolation theory, among other things, 
aims at studying the behavior of the volume of the 
largest cluster 
as a function of the density threshold or equivalently the filling
factor $(FF)$. 
At a given density threshold $\delta$ a cluster
is identified as a connected overdense region, connectivity being defined
using nearest neighbors. In Fig.~\ref {perc} we have shown the results of 
percolation analysis of density fields obtained for the models
described above;  the scale of nonlinearity increases 
from top to bottom.
The ratio of the fraction of volume in the largest cluster to the fraction of
volume in all clusters (i.e. $FF$), denoted
$v_\infty,$ is shown plotted against $FF$ (thick solid line). The ratio 
of the total volume of all but the largest cluster to  $FF$ of 
overdense regions, denoted $v_0,$ is also shown (thick dashed line). 
Error bars represent the dispersion in $v_\infty$  and $v_0$ 
computed with the aid of four different
realizations of a given spectrum of fluctuations. The thin vertical
line in each panel corresponds to the {\it critical} $FF = FF_c$ 
(or $\delta =\delta_c$)
at which the percolation
transition formally sets in (i.e. the opposite faces of the cube are linked);
the region to the right of the line corresponds to
the phase in which the largest cluster percolates and the region to the
left corresponds to the phase in which it does not.
It is clearly seen that $FF_c$ is smaller (and the associated
density threshold higher) for the $n=-2$ model 
relative to the $n=0$ model. 
This may be related to the fact that the $n=-2$ model, 
has considerable large scale power making percolation easier in this case.
With the onset of nonlinearity on larger scales
(i.e. during later epochs) $FF_c$ monotonically decreases for both models
indicating the increasing coherence of structure
as larger scales go nonlinear. 
The dispersion in the curves being small the features
mentioned above are statistically significant. Thus we find that
percolation analysis provides a sensitive probe of the 
nature of gravitational collapse in different
scenarios of clustering (see also Yess \& Shandarin 1996). 

At a value slightly lower than $FF_c$ the volume in all clusters 
except the largest one reaches a maximum
(Yess \& Shandarin 1996).
This maximum
is generic and occurs just before criticality irrespective
of the spectrum of fluctuations and the stage of nonlinearity. 
At $FF > FF_c$ many small clusters link up
with the largest cluster thereby aiding percolation, on the other hand 
at $FF < FF_c,$ clusters tend to be sparse and gradually disappear as
$FF \rightarrow 0$. Both these trends imply that the cluster 
distribution should peak before criticality which it does. 
We speculate
that in an infinite, continuous system the occurrence of this 
maximum will coincide with $FF_c$; however the size and resolution of the
present simulations are insufficient to unambiguously settle this 
issue which will be taken up in a future work. Thus, the percolation
transition provides an objective choice of threshold at which to identify
clusters. 
The results presented in this Letter concerning the morphology
of structure are obtained by studying overdense regions slightly
{\it below} the critical threshold
in terms of the filling factor: $\sim FF_c$ ({\it above} the critical
threshold in terms of the density contrast: $\delta > \delta_c$).

Babul and Starkman (1992)
characterize the morphology of structure in a
region of radius $R$ around a given point using a triad of numbers which
they call {\it structure functions.}  They begin with the first and 
second moments of the distribution of particles around a fiducial
point. With the aid of these moments they construct the moment of
inertia tensor and its associated eigenvalues.
They then invoke a nonlinear transformation
to arrive at three structure functions $S_1,$ $S_2$ and $S_3$ each
of which takes on values in the range [0,1] and exhibits unbiased
behavior as a spherical distribution is continuously deformed either into
a sheet or a filament, and a sheet is deformed into a filament (and vice-versa)
(see [\cite{bs92,sss95}] for more details).
The triad of structure functions is given by
\begin {equation}
S_1 \equiv \sin \left [ {\pi \over 2} \left (1-\mu\right )^p \right ]
\end {equation}
\begin {equation}
S_2 \equiv \sin \left [ {\pi \over 2}  a(\mu,\nu)\right ]
\end {equation}
\begin {equation}
S_3 \equiv \sin \left [ {\pi \nu \over 2} \right ]
\end {equation}
where $\mu\equiv \sqrt {\lambda_2/\lambda_1},$  
$\nu\equiv \sqrt {\lambda_3/\lambda_1}$ ($\lambda_1,$  $\lambda_2,$  and
$\lambda_3$ being the three eigenvalues of the moment of inertia tensor arranged
in decreasing order), the function $a(\mu,\nu)$ is implicitly defined by
the equation 
\begin {equation}
{\mu^2 \over a^2} - {\nu^2 \over a^2 \left (1-\alpha a^{1/3} +
\beta a^{2/3}\right )} = 1.
\end {equation}
The values of parameters $p=\log 3/\log 1.5$, $\alpha=1.9$ and 
$\beta=-(7/8)~9^{1/3} + \alpha~3^{1/3}$, are so chosen to normalize $S_n$.
It is both instructive and legitimate to visualize the triad
of structure functions as a vector ${\bf S}=(S_1,S_2,S_3)$ lying in
the first octant of the three-dimensional space with magnitude
$|{\bf S}|\le 1,$ the x-axis (say) characterizing linearity
(filamentarity) $S_1$, the y-axis corresponding to planarity 
(oblateness) $S_2$, and the z-axis representing sphericity $S_3,$ of the
distribution of particles/points around a fiducial point. A vector
lying along one of the x-, y- or z-axis is of unit magnitude by construction
and represents, respectively, a perfectly linear, planar or spherically
symmetric distribution of particles around the point in question. 
We shall however consider
density fields defined on a grid, and {\it not} point processes, and use 
appropriately modified density weighted moments of the distribution. 
Such a modification
overcomes the need to have particle positions and enhances the scope
of the shape statistics as it can now be applied to study
morphology of scalar fields in a variety of different contexts. 
(The quantity defined on the
grid could be as diverse as: the surface brightness of a galaxy,
the temperature of the sky as measured by a microwave background
experiment, etc. It could also  be a binary state taking on values 0/1.)

Conventionally the behavior of shape statistics has been studied as a function
of the size of the region by averaging structure functions over a large
ensemble of randomly chosen points (for 
an alternate application involving Minimal Spanning trees see [\cite{pc95}]). 
In addition to measuring
the average value of the shape statistic for the entire ensemble 
we also study the shape of an
individual cluster by choosing a window large enough to enclose the
entire cluster and by setting the density at all but points on the
cluster to zero. The origin about which the moments of the density
field are computed is the center of mass of the cluster.

In Fig.~\ref {hist1} we have shown the histogram of the three structure
functions of isolated clusters, computed as described above, and plotted
as a function
of the cluster mass. In all the panels linearity is shown as a thick
solid line, planarity as a dotted line and sphericity as a dashed line.
The height of the histogram on a given mass scale represents the average
value of the corresponding statistic obtained using an ensemble of
four distinct realizations of a single power spectrum, 
error bars are 1$\sigma$ dispersions over the ensemble. As in Fig.~\ref {perc}
left panels correspond to $n=-2$ models and right panels to $n=0$ models,
the scale of nonlinearity increases from top to bottom.
In both spectra at earlier epochs the statistics suggests that 
clusters are predominantly spherical irrespective of their mass.
With the onset of nonlinearity on larger scales,
first low mass clusters begin to develop filamentarity and, at 
later epochs, larger clusters too become filamentary
(this effect is particularly noticeable for $n = -2$). 
Except on the very largest
scales (and during some epochs) filamentarity dominates over planarity, the
difference between the two being statistically significant.

For the largest structures the measure of sphericity must
be treated with some caution. In particular it is worth stressing that 
when measured on the largest scales, the statistics 
we use {\it cannot} distinguish between a web
of connected filaments bordered by a large sphere from a smooth mass 
distribution in a similar sphere.

In Fig.~\ref {avg} we plot structure functions averaged over many random
but high density points as functions of the window size $R.$ Curves
are as in Fig.~\ref {hist1} except that here we show 
four epochs in the same panel with heavier lines
representing later stages of nonlinearity. The average value of 
shape functions spectacularly demonstrates our claim concerning the
emergence of filamentarity as gravitational clustering proceeds from
the mildly to the strongly nonlinear regime. We see that structures
tend to become oblate {\it only after} a filament-like shape has already
developed. We find that (except on the largest scale)
prolateness dominates over oblateness
at all epochs for both spectra.
It is likely that at later epochs 
neighboring clusters align themselves leading to the increasing amplitude of
curves representing filamentarity in  Fig.~\ref {avg}. This
behavior is seen for planarity too although at a smaller amplitude.

We conclude our analysis of:
(i) shapes of isolated clusters and 
(ii) shapes of overdense regions averaging all
clusters, by emphasising that
filamentarity 
dominates over planarity for both (i) and (ii), growing
more pronounced at later epochs. Although almost always weaker than 
filamentarity, planarity too is statistically significant. Thus, the mass
distribution in a generic cosmological scenario of gravitational 
clustering is in a good {\it qualitative} agreement with 
the galaxy distribution
characterized as ``a network of surfaces'' by de Lapparent et al (1991).

It is worth stressing that the visual dominance of filaments over 
planar structures 
in the density distribution is typical even for classical pancake
models such as the HDM scenario (\cite{ks83}). The underlying 
generic singularities (to be realized at later times)
other than pancakes were suggested to explain qualitatively this phenomena 
(\cite{asz82}). Thus, the results of this
paper do not
contradict the conclusion that the first singualrities are pancakelike
(\cite{shetal95}).

Since filamentarity increases with decreasing spectral index $n$, we expect
to see more filamentary objects at higher redshifts in CDM-like models
for which the effective index shows a significant tilt towards
smaller scales.

Clearly an important issue not resolved in the present paper, concerns the
{\it quantitative} morphology and shape analysis 
of three dimensional galaxy distributions. We shall 
address this issue in a subsequent paper (\cite{sssk95}),
where we evaluate structure functions for the IRAS density map.

\bigskip
\noindent {\bf Acknowledgments:}
For useful conversations and critical remarks we thank Tarun Souradeep Ghosh,
Dipak Munshi, Dimitry Pogosyan, Lev Kofman and Adrian Melott. 
Acknowledgments are also due to the Smithsonian Institution, Washington,
USA, for International travel assistance under the ongoing Indo-US
exchange program at IUCAA, Pune, India.
One of us (BSS) would like to thank the Department of Physics and Astronomy,
University of Kansas at Lawrence for hospitality. S. Shandarin acknowledges
NSF grant AST-9021414, NASA grant NAGW-3832, and the University of Kansas
GRF-96 grant.

\begin {figure}
\caption {Percolation transition in $n=-2$ model (left panels) and $n=0$ model
(right panels) at different epochs measured by the scale of 
nonlinearity $k_{\rm NL}^{-1}$ which increases from top to bottom. 
The filling factor of the
largest cluster $v_\infty$ (solid line) and the filling factor of all but the 
largest cluster $v_0$ (dashed line) (both measured in units of the total filling
factor) are plotted against the total filling factor $FF.$}
\label {perc}
\end {figure}

\begin {figure}
\caption {The average linearity (solid lines), planarity (dotted lines) 
and sphericity (dashed lines) of clusters belonging to
a given mass range are plotted
as functions of cluster mass. As in Fig.~1 the scale of nonlinearity 
increases from top to bottom.}
\label {hist1}
\end {figure}

\begin {figure}
\caption {Structure functions averaged over random (but high density)
points are shown as functions of the window radius $R$ (measured in units of the
grid size). Curves are for the epochs $k_{\rm NL} = 32, 16, $ 8 and 4
(in units of the fundamental mode) with heavier lines corresponding to
later epochs.  }
\label {avg}
\end {figure}


\begin{thebibliography}{}
\bibitem [Arnol'd et al. 1982]{asz82}
Arnol'd, V.I., Shandarin, S.F., \& Zel'dovich, Ya.B. 1982, Geophys. Astrophys.
Fluid Dynamics, 20, 111
\bibitem [Babul \& Starkman 1992] {bs92}
Babul A., \& Starkman G.D. 1992, ApJ, 401, 28
\bibitem [Dominik \& Shandarin 1992] {ds91}
Dominik K., \& Shandarin S.F. 1992, ApJ, 393, 450
\bibitem [Klypin \& Shandarin 1983] {ks83}
Klypin A.A., \& Shandarin S.F. 1983, MNRAS, 204, 891
\bibitem [Klypin \& Shandarin 1993] {ks93}
Klypin A.A., \& Shandarin S.F. 1993, ApJ, 413, 48
\bibitem [de Lapparent et al. 1991]{delgh91}
de Lapparent, V., Geller, M.J., \& Huchra, J.P. 1991, ApJ, 369, 273
\bibitem [Melott \& Shandarin 1993]{msh93}
Melott, A.L, \& Shandarin, S.F. 1993, ApJ, 410, 469
\bibitem [Pearson \& Coles 1995] {pc95}
Pearson R.C., \& Coles P. 1995, MNRAS, 272, 231
\bibitem [Sahni \& Coles 1995] {sc95}
Sahni V., \& Coles P., 1995, Physics Reports, 262, 1
\bibitem [Sathyaprakash et al. 1995a] {sss95}
Sathyaprakash B.S., Sahni V., \& Shandarin S.F. 1995a, MNRAS, submitted
\bibitem [Sathyaprakash et al. 1995b] {sssk95} 
Sathyaprakash B.S., Sahni V., Shandarin S.F., \& Fisher K. 1995b, 
in preparation
\bibitem [Shandarin 1983]{sh83}
Shandarin, S.F. 1983,  Soviet Astron. Lett., 9, 104
\bibitem [Shandarin \& Zel'dovich 1983]{shz83}
Shandarin, S.F., \& Zel'dovich Ya. B. 1983, Comments Astrophys., 10, 33
\bibitem [Shandarin et al. 1995]{shetal95}
Shandarin, S.F., Melott, A.L., McDavitt, K., Pauls, J.L., \& Tinker, J.
1995, Phys. Rev. Lett., 75, 7
\bibitem [Yess \& Shandarin 1996]{ys96}
Yess C., Shandarin S.F., 1996, ApJ, to be published
\bibitem [Zel'dovich 1982] {zel82}
Zel'dovich Ya. B., 1982, Soviet Astron. Lett., 8, 102
\end{thebibliography}
\end {document}
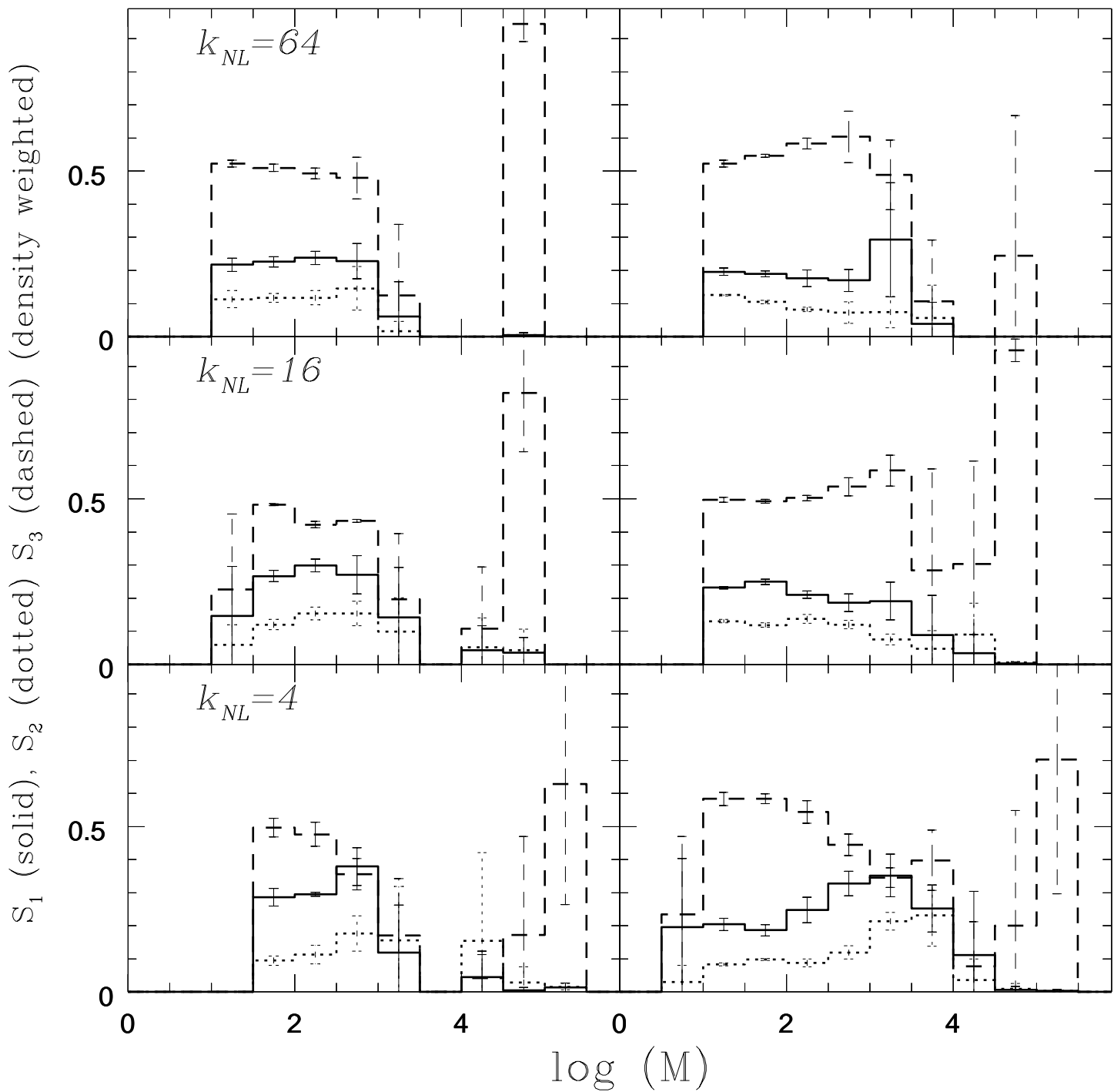
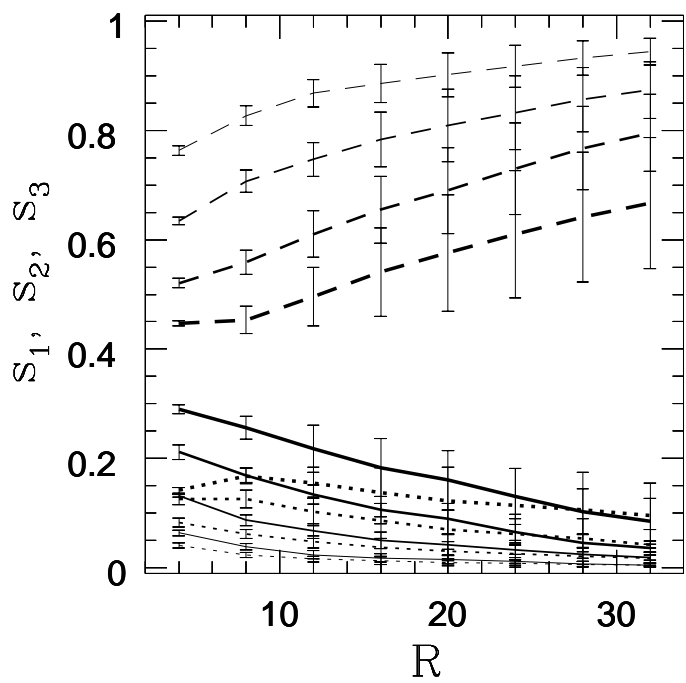
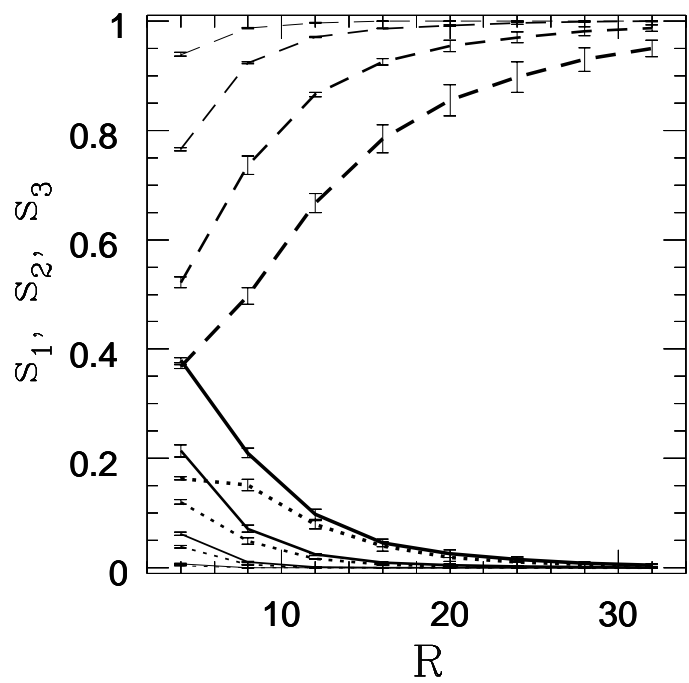
